\documentclass{aa}
\usepackage{graphics}
\usepackage{longtable}
\begin{document}
\title{Fe~II lifetimes and transition probabilities}
\author{R. Schnabel
          \and
          M. Schultz-Johanning
          \and
          M. Kock
          }

\offprints{R. Schnabel}
\institute{Institut f\"ur Atom- und Molek\"ulphysik, Abteilung Plasmaphysik, \\
Universit\"at Hannover, Callinstrasse 38, D-30167 Hannover, Germany \\
email: roman.schnabel@aei.mpg.de}

\date{Received }
\titlerunning{Fe~II lifetimes and transition probabilities}
\authorrunning{Schnabel et al.}

\abstract{Fe~II radiative lifetimes were measured applying the
time-resolved nonlinear laser-induced fluoresence technique. We
investigated 21 levels of up to 47000 cm$^{-1}$. The uncertainties
are typically 2~--~3~\%. The lifetimes provide an improved
absolute scale to our branching fractions which were measured with
a Fourier transform spectrometer and a high--resolution grating
spectrometer and which have been published earlier. We report
absolute transition probabilities of 140 Fe~II lines in the
wavelength range 220--780 nm. The overall uncertainties are
estimated to be 6~\% for the strong and up to 26~\% for the weak
transitions. The results are compared with recent experimental
data from the literature.
Our large set of accurate data can be used for a reliability check
of theoretical data calculated for iron abundances in
astrophysical plasmas.}


\maketitle

\section{Introduction}

We developed an improved apparatus to measure lifetimes below 5~ns
more accurately with the time-resolved laser-induced fluorescence
(LIF) technique. The improvements include a more sophisticated
evaluation procedure accounting for saturation in the LIF signals.
Additionally a linear radio-frequency ion trap (Paul trap) was
designed to improve the signal--to--noise ratio in the lifetime
measurements. Up to now the most accurate lifetime measurements in
Fe~II were performed using the fast beam--laser technique
(\cite{BBKAP91}, \cite{GAPJB92}). Our improvements to the
time-resolved LIF technique now enable a critical comparison of
experimental lifetimes derived from
both techniques.\\
A radiative lifetime provides an absolute scale to the branching
fractions of that level. The result of this combination is a set
of absolute transition probabilities (\cite{HSa86} and
\cite{Koc96}). In \cite{Kroll85}, \cite{KKo87} and \cite{HKo90} we
used lifetimes with high uncertainties of 10~\% to calibrate our
experimental branching fractions in the Fe~II spectrum. These
lifetimes were also affected by a systematic error, since they
proved to be systematically too large. The recent improvements in
lifetime measurements enable us to report a much more reliable set
of Fe~II transition probabilities of 140 Fe~II lines in the
wavelength range 220--780 nm. 13 of these lines have already been
used for a solar iron abundance determination (\cite{SKH99}).

\section{Lifetime Measurement}

The time-resolved laser-induced fluorescence (TRLIF) is a standard
technique for measuring radiative lifetimes of excited atomic
states.
The excitation is done with short laser pulses, ideally much
shorter than the radiative lifetime under investigation. The LIF
signal is recorded over a time scale more than ten times of the
radiative lifetime by a fast photo-multiplier in conjunction with
a digitizing oscilloscope. Provided that a sufficiently large
number of atoms is available, a complete decay curve can be
obtained from a single laser pulse. TRLIF therefore is a
single--shot method.

In recent papers (\cite{SKo00a}, \cite{SKo00b}) we examined in
detail the TRLIF method by including for the first time saturation
in laser excitation quantitatively. This was done by using a rate
equation model for a two-level and for a three-level atom,
respectively. We were able to show that a natural lifetime, which
is shorter than the laser pulse, can still be measured with an
uncertainty of a few percent. Systematic error sources due to
saturation or a non--ideal response of the detection system could
be eliminated.

The rate equation model is adequate as long as the atoms are not
coherently excited. In this regard we have to consider vertical
coherences (Rabi oscillations) as well as horizontal coherences
(polarization, alignment of sublevels). Overlapping modes in the
multimode laser beam produce a broadband excitation and coherence
times shorter than the relevant nanosecond time scale of our
experiment. Horizontal coherences are produced by the anisotropic
excitation of the polarized laser light. In the present experiment
we used low saturation parameters, i. e. weak laser pumping,
although saturation is hard to avoid completely without severe
loss in signal-to-noise ratio. With low saturation observation of
LIF under the angle of 54.7$^\circ$ with respect to the axis of
laser polarization (magic--angle arrangement) ensured a
time-resolved LIF signal comparable to isotropic excitation
experiments (\cite{HLo83}). However, we could not find any
anisotropy in the LIF signals when varying the polarization axis
of the laser, indicating an internal depolarization mechanism of
the trapped ions. This has been confirmed by experiments as
discussed by \cite{SKo00b}.

As a particle source, we used a high--current hollow cathode
combined with a linear Paul trap. This apparatus was designed and
optimized for pulsed laser spectroscopy on large numbers of metal
ions in their ground state or in metastable states (\cite{LSK98}).

The Paul trap was operated as an ion guide. Whereas the neutral
particles expanded effusively, ionized species were collimated
onto the axis due to their charge/mass ratio. The ion trap/guide
arrangement consisted of four copper cylindrical rods as guiding
electrodes. The thermal population of metastable states inside the
discharge was conserved and exploited successfully (\cite{SSK99}).

Two different laser systems were at our disposal. At the Lund
Laser Centre (LLC), stimulated Brillouin scattering (SBS) was used
to generate short laser pulses below 1~ns (\cite{LNPWSDB99},
\cite{Norin98}). Compressed second--harmonic pulses of an
injection--seeded Nd:YAG laser were used to pump a dye laser. In
the present experiment we obtained linearly polarized pulses of
0.5~ns duration (FWHM) with a repetition rate of 10~Hz in the
wavelength range 210-270 nm. At our department longer lasting
laser pulses of 2 to 3~ns ns duration, also linearly polarized and
with a repetition rate of 10~Hz, are used. The third harmonic of
our Quanta Ray DCR 11--3 Nd:YAG laser pumps a Lambda Physik LPD
3002 dye laser. Overlapping laser modes provided a broadband
excitation of the Doppler--broadened ionic lines. We estimated the
transition line width of the cooled ions inside the trap to be
smaller than 2 GHz.

The Fe~II ions in the collimated beam were excited resonantly at a
distance of 4~cm or 15~cm, respectively, measured from the small
exit aperture. Perpendicular to the laser beam the fluorescence
photons were imaged by a lens system onto the photo detector. The
crossing of atom beam and laser beam was located in the center of
a Helmholtz coil. Magnetic fields of up to 2~mT could be
superimposed to check the influence of quantum beats. Also the
polarization axis of the laser was varied to investigate
anisotropy in the LIF signals. Both effects could not be observed
in this experiments on Fe~II, although observed previously on
different species (cf. \cite{SKo97} and \cite{SKo00a}).

Fluorescence photons were recorded time-resolved using a fast
photo detector in combination with a fast digitizing oscilloscope
(Tektronix TDS 680~B). We used two different photo detectors. At
LLC a Hamamatsu 1564U micro-channel plate (MCP) with a rise-time
of 200~ps was at our disposal. Additionally we used a Hamamatsu
R2496 photomultiplier with a rise-time of 700~ps. In both cases
the response function of the detector system, including signal
cable and oscilloscope was measured separately and entered our
evaluation procedure \cite{SKo00a}. Our oscilloscope had an analog
bandwidth of 1~GHz and a real--time scanning rate of 2 $\times$
5~Gigasamples/s.

The {\sl time-resolved nonlinear LIF} technique requires recording
of the temporal and spatial intensity distributions of the laser
pulses. The temporal shape was measured with a fast photo tube
(Hamamatsu R1328U--52) simultaneously with the measurement of the
LIF signal using the second channel of the oscilloscope. This
signal was also used to trigger the fluorescence measurement. The
rise-time of that detection system was (330 $\pm$ 20)~ps measured
with femtosecond laser pulses and was mainly limited by the
oscilloscope. The spatial photon distribution of the laser pulses
was measured with a 2/3~inch {\sl CCIR Standard} CCD chip with 756
$\times$ 583 pixels of 11~$\mu$m size. For each single laser pulse
an 8--bit histogram of the intensity was recorded. This procedure
was fast enough to handle the 10~Hz repetition rate of our
experiment. For the evaluation procedure an averaged 15--bit
histogram was calculated from the single shots. The temporal
fluorescence was also averaged over the single shots. A typical
single lifetime measurement lasted over 100~s (1000 laser pulses).
The measured averaged spatial and temporal laser pulse profiles
were used to solve the rate equations for a three-level atom. The
result was a temporal process of expected fluorescence intensity
which was convoluted with the response function of the detection
system and then fitted to the measured signal. For the result we
derived the level lifetime which was one of four fitting
parameters. For more detailed information we refer to
\cite{SKo00a} and \cite{SKo00b}.

\section{Branching Fraction Measurement}

In the present paper we use branching fractions from previous
publications, the PhD thesis of \cite{Kroll85}, and the journal
papers by \cite{KKo87} and \cite{HKo90}. In all cases branching
fractions were determined by means of emission spectroscopy.
Additionally Kroll used dispersion (hook) measurements to check
the emission measurements to be free from systematic errors. The
hook measurements also allowed a consistency check of lifetime
data, thereby allowing an optimization of lifetime values which
were available from literature. For all emission measurements we
used our large-scale hollow cathode lamps. The iron spectra were
recorded by our 2\,m McPherson monochromator with a plane grating
of 2400 lines per mm and by the 1\,m Fourier-transform
spectrometer at Kitt Peak National Observatory in Tucson, Arizona.
A detailed description of the experimental procedures used can be
found in the three references given above.

\section{Results and Discussion}

As already said above, absolute transition probabilities are
derived from independent measurements of lifetimes and full sets
of branching fractions. Both experimental procedures involve
different techniques and are subject to systematic and statistical
uncertainties of different size. In the following paragraphs we
therefore discuss our lifetime and branching fraction data
separately.

Table~1 summarizes the results of our lifetime measurements. Here
error bars are given in absolute values and represent a
$1\,\sigma$ standard deviation derived from up to 20 independent
lifetime measurements. Systematic errors have been estimated to be
negligible. Comparing our set with accurate literature data we
find excellent agreement confirming the absence of systematic
error sources in both measuring techniques, the fast beam--laser
technique (\cite{BBKAP91}, \cite{GAPJB92}) and the time-resolved
non-linear LIF technique used in this work. Since we used an ion
guide our work was not effected by possible blends and
misidentifications due to Fe~I lines. Misidentifications due to
Fe~II lines were excluded by using more than one excitation
wavelength for the upper levels. This was done for more than half
of the levels giving us a sufficient absolute wavelength
calibration for all the laser dyes used.

In Table~2 we present our branching fractions which are
re-separated from the absolute data given in \cite{Kroll85},
\cite{KKo87} and \cite{HKo90}. Possible transitions which were not
found in the spectra have been estimated with the data of
\cite{KBe95} to contribute less than $2\%$ and have been accounted
for in Table~2. We compare our branching fractions with
experimental data from \cite{BMWLLJ96}. In the last row of Table~2
our lifetime data is used for a new absolute scaling of our
experimental branching fractions. For two levels, lifetime values
have been taken from literature.

In Table~2 all error bars are given in $\%$. The error bars of our
data include statistical and estimated systematic uncertainties.
For the transitions at 430.317~nm and 523.462~nm we found
different results in our investigations, here we quote the values
as given in \cite{HKo90}. For a quantitative comparison with data
from \cite{BMWLLJ96}, we quote their error bars including a 5~\%
calibration uncertainty as given in that paper. For virtually
$90\%$ of all the lines measured, a good common agreement within
the mutual error bars has been found. The comparison shows no
runaways, however, the sets of Bergeson {\it{et al.}} are far from
complete.

Considering now the absolute data in Table~2, we find our new
transition probabilities to be increased by about 10$\%$ on
average compared with our values published earlier. This
corresponds to the new lifetime values applied. The uncertainties
in our absolute data are slightly reduced due to the higher
accuracy of the lifetimes. Although not explicitly given in
Table~2, we also compared the absolute transition probabilities.
Again good agreement is found for the cases where accurate data
are available from literature (\cite{BMWLLJ96}). Comparing our
absolute data with the subset we used for a solar iron abundance
determination in \cite{SKH99}, we find 4 values to be decreased by
about 4\,\% and 4 other values to be increased by 2\,\%. The
reasons are slightly changed estimated contributions by unknown
lines. Altogether there is no effect on our published solar iron
abundance of logN(Fe)$=7.42 \pm 0.09$.

Our data sets are suitable for a reliability check of theoretical
data. We refer to theoretical data that can be found in
\cite{KBe95}, \cite{RUy98b} and \cite{RUy98a}. The consistency
with our data is generally quite good although the comparison is
not provided in Table~2. In cases where no experimental data is
available yet it is generally quite difficult to estimate the
quality of theoretical data. The completeness and the accuracy of
our data might help theorists to derive such an estimation, taking
into account the details of their theoretical codes and the
experimental data used for semi-empirical approaches.

In conclusion, we have presented 140 Fe~II transition
probabilities. Given the high degree of completeness of the sets
and the accuracy of the lifetime values used, Table~2 presents one
of the most reliable values at present. A substantial improvement
in accuracy can only be achieved in the branching fraction data.
Although an independent technique to measure strong branches has
been demonstrated recently (\cite{SKo00b}) it remains challenging
to reduce systematic and statistical uncertainties in branching
fraction measurements.

\begin{figure*}
\begin{center}
\resizebox{8.5cm}{!}{\includegraphics{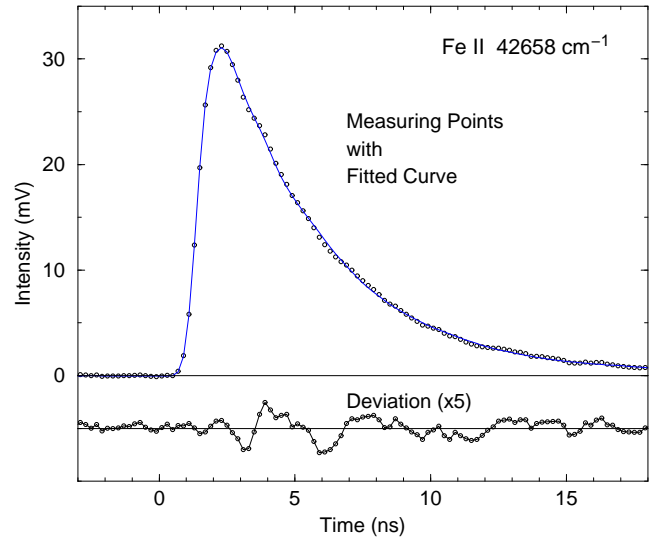}}\\
\vspace{3mm}
\end{center}
\caption{Time-resolved nonlinear LIF signal during and after a
short laser pulse excitation. The circles represent the
measurement and the solid line is the fitted theoretical curve.
From several of such measurements we determined the lifetime of
this Fe~II level to 3.71~ns with a standard deviation of $\pm$
0.04~ns.}
\end{figure*}

\begin{figure*}
\begin{center}
\resizebox{14.5cm}{!}{\includegraphics{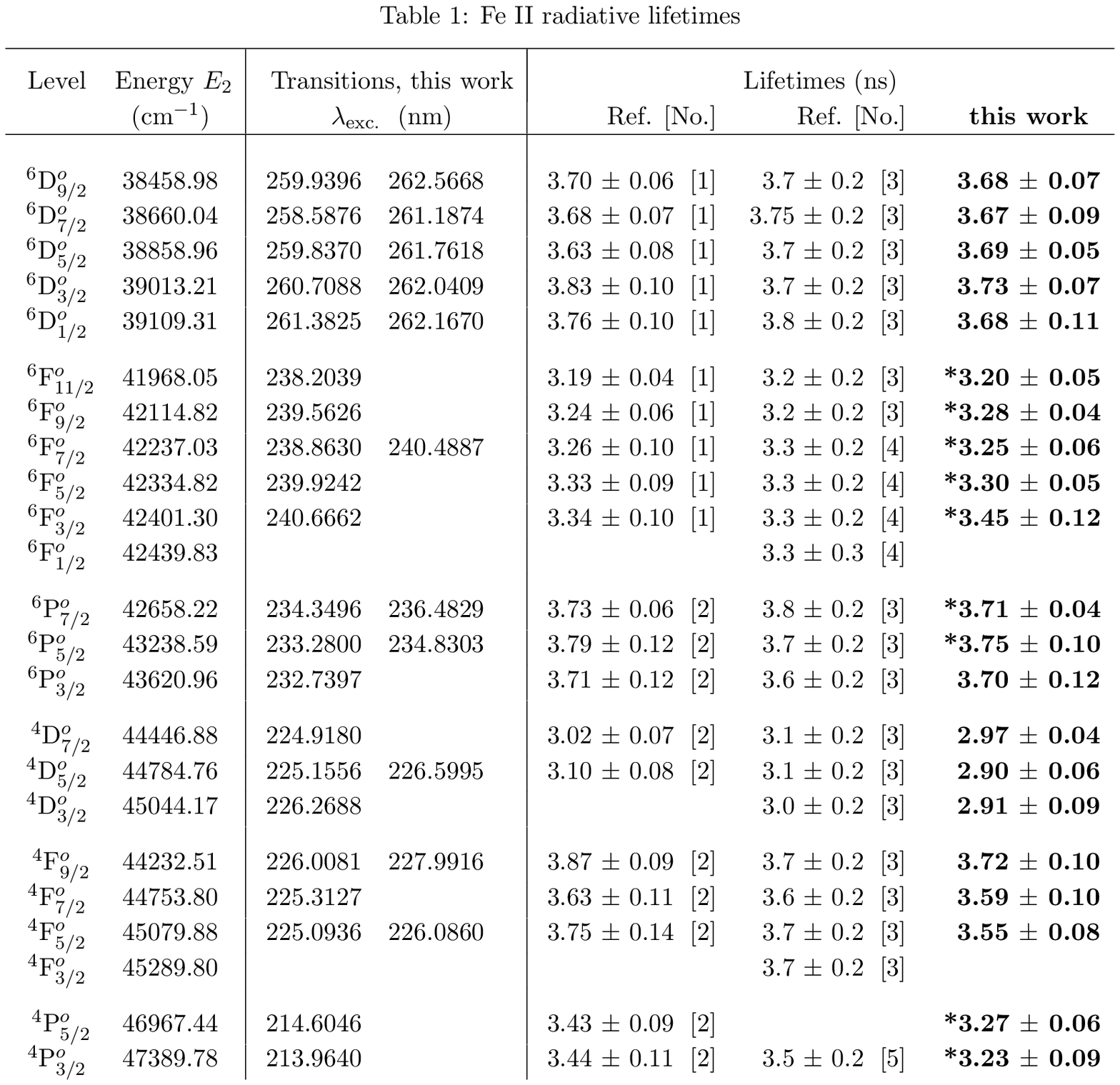}}\\
\vspace{3mm}
{[1]~\cite{BBKAP91}, $\,\,$ [2]~\cite{GAPJB92}, $\,\,$ [3]~\cite{HLGN92},
$\,\,$ [4]~\cite{SMH88},}\\ {$\,\,$  [5]~\cite{LLSJ99}.\rule{116mm}{0mm}.\\[2mm]
* Lifetime was measured with SBS--compressed laser pulses.\rule{52mm}{0mm}}
\end{center}
\end{figure*}

\begin{figure*}
\begin{center}
\resizebox{14.5cm}{!}{\includegraphics{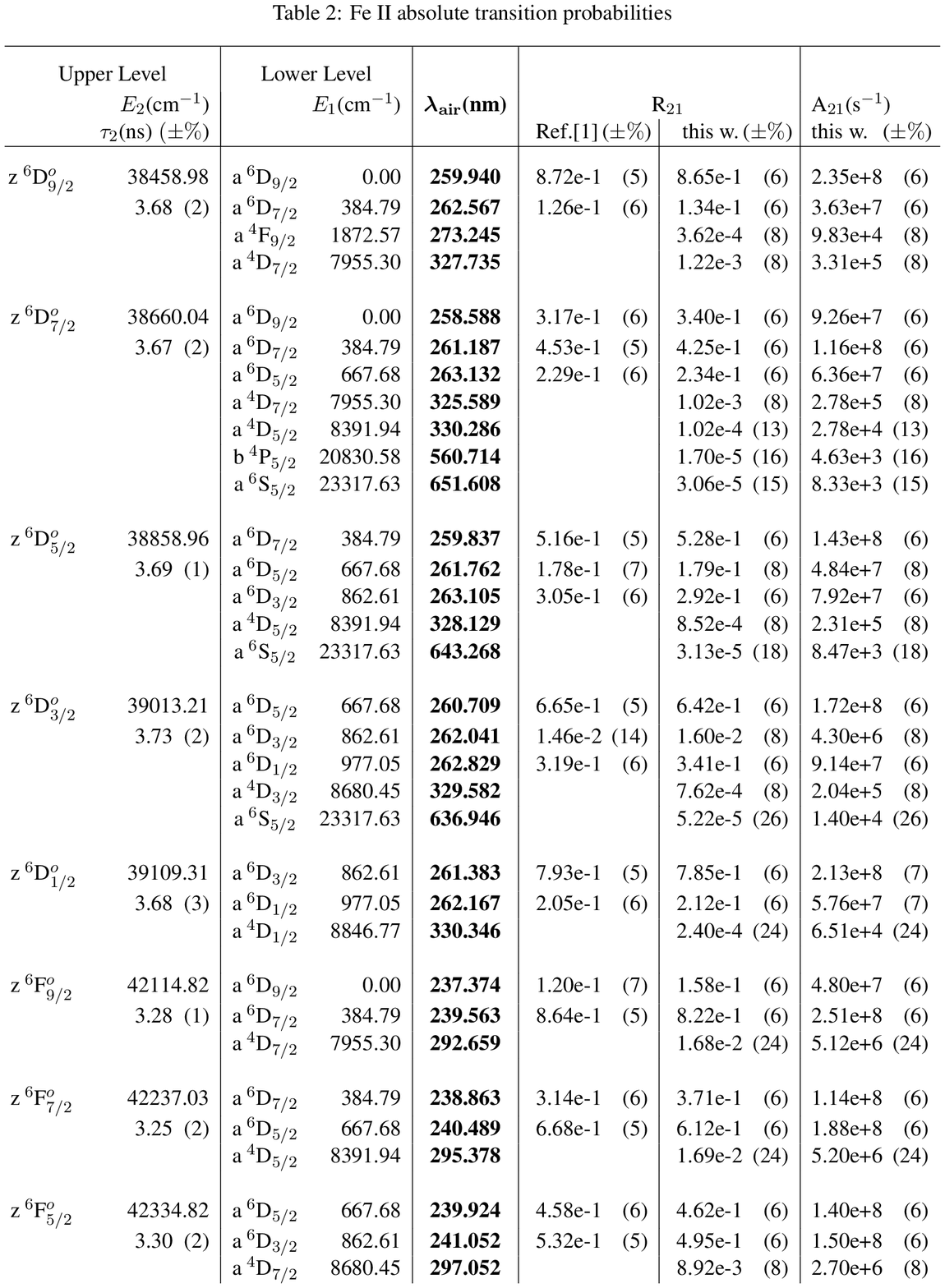}}
\vspace{3mm}\\
{[1]~\cite{BMWLLJ96}.}\\
\end{center}
\end{figure*}

\begin{figure*}
\begin{center}
\resizebox{14.5cm}{!}{\includegraphics{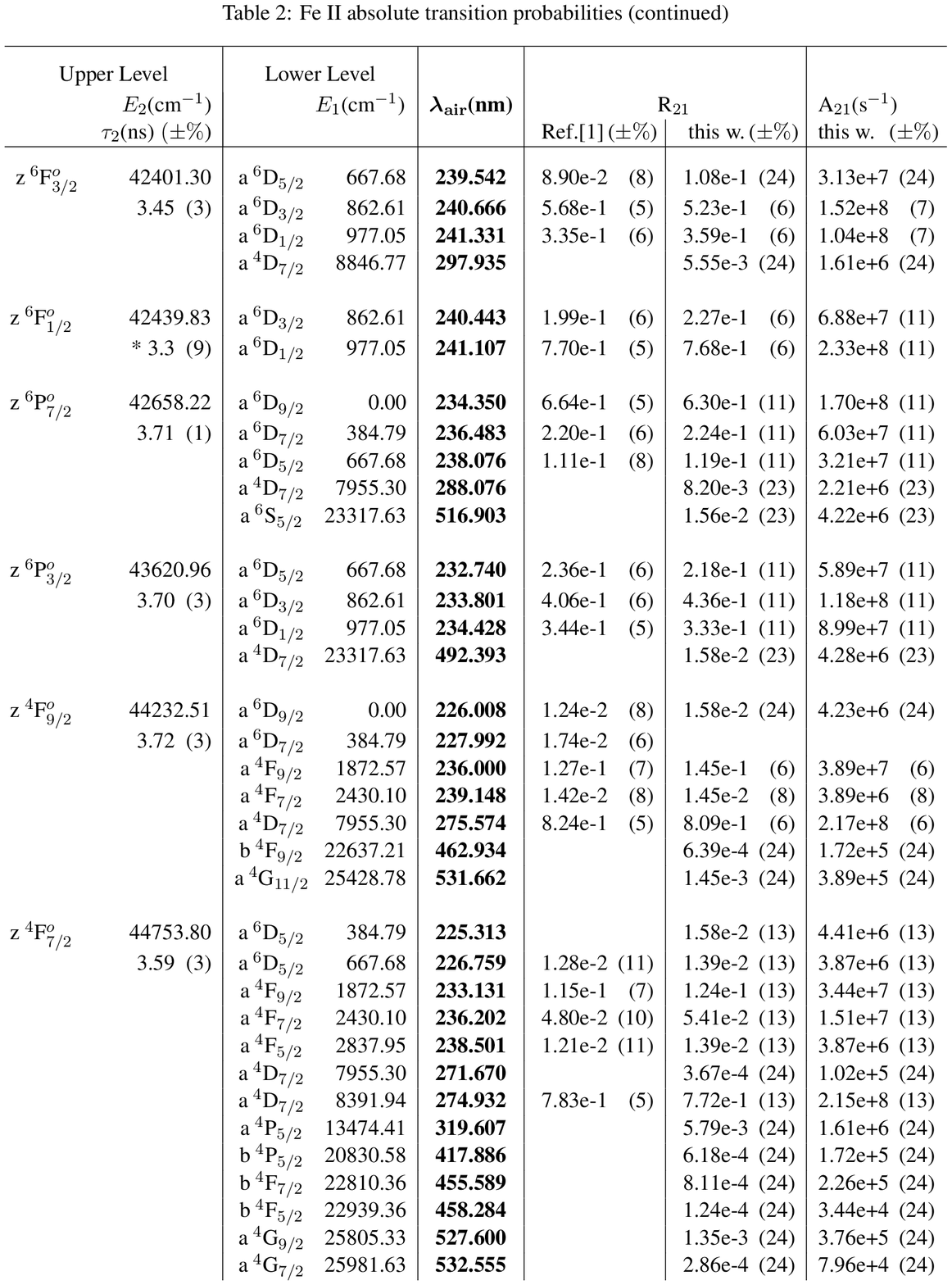}}
\vspace{3mm}\\
{[1]~\cite{BMWLLJ96}.}\\
{* Lifetime taken from \cite{SMH88}.}
\end{center}
\end{figure*}

\begin{figure*}
\begin{center}
\resizebox{14.5cm}{!}{\includegraphics{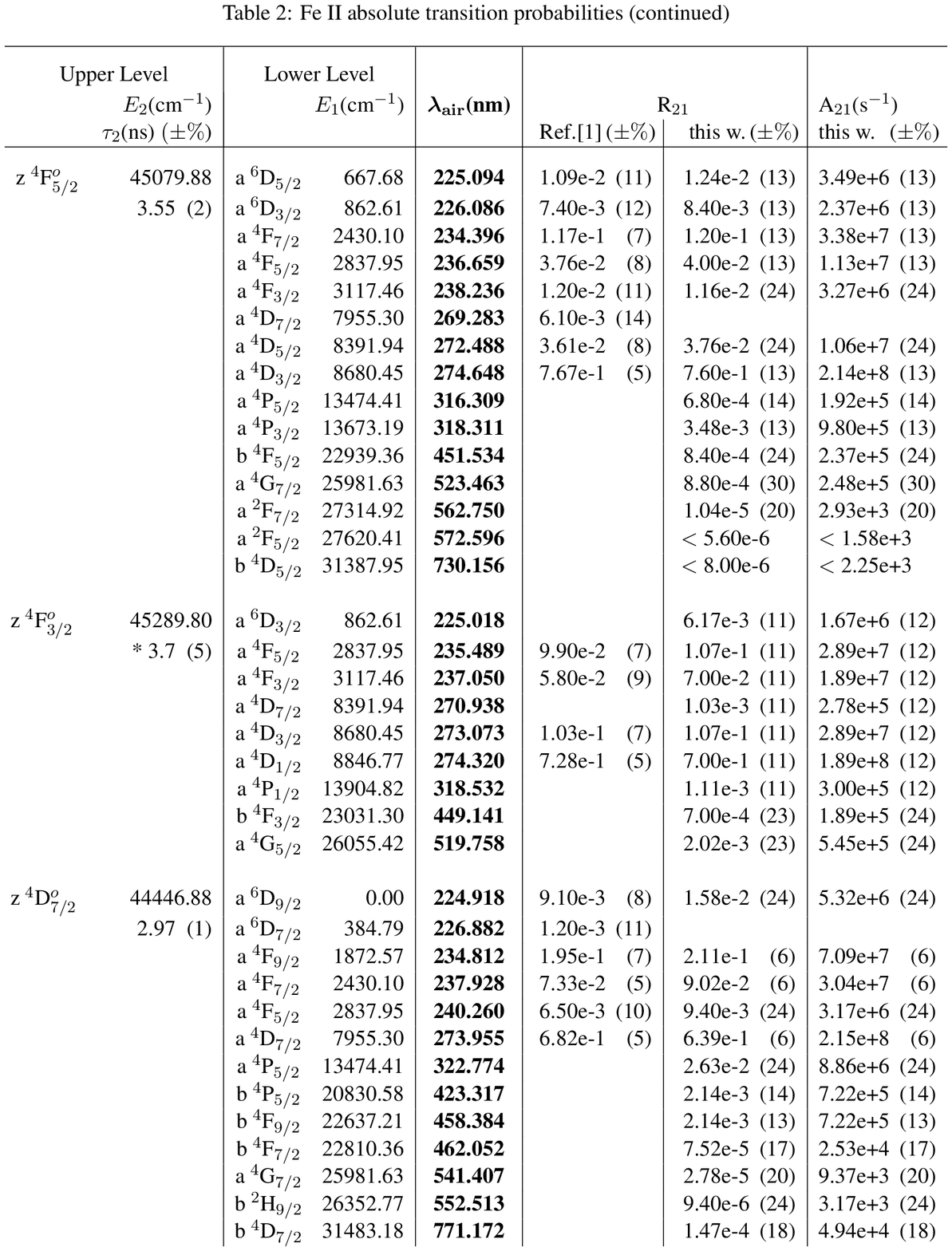}}
\vspace{3mm}\\
{[1]~\cite{BMWLLJ96}.}\\
{* Lifetime taken from \cite{HLGN92}.}
\end{center}
\end{figure*}

\begin{figure*}
\begin{center}
\resizebox{14.5cm}{!}{\includegraphics{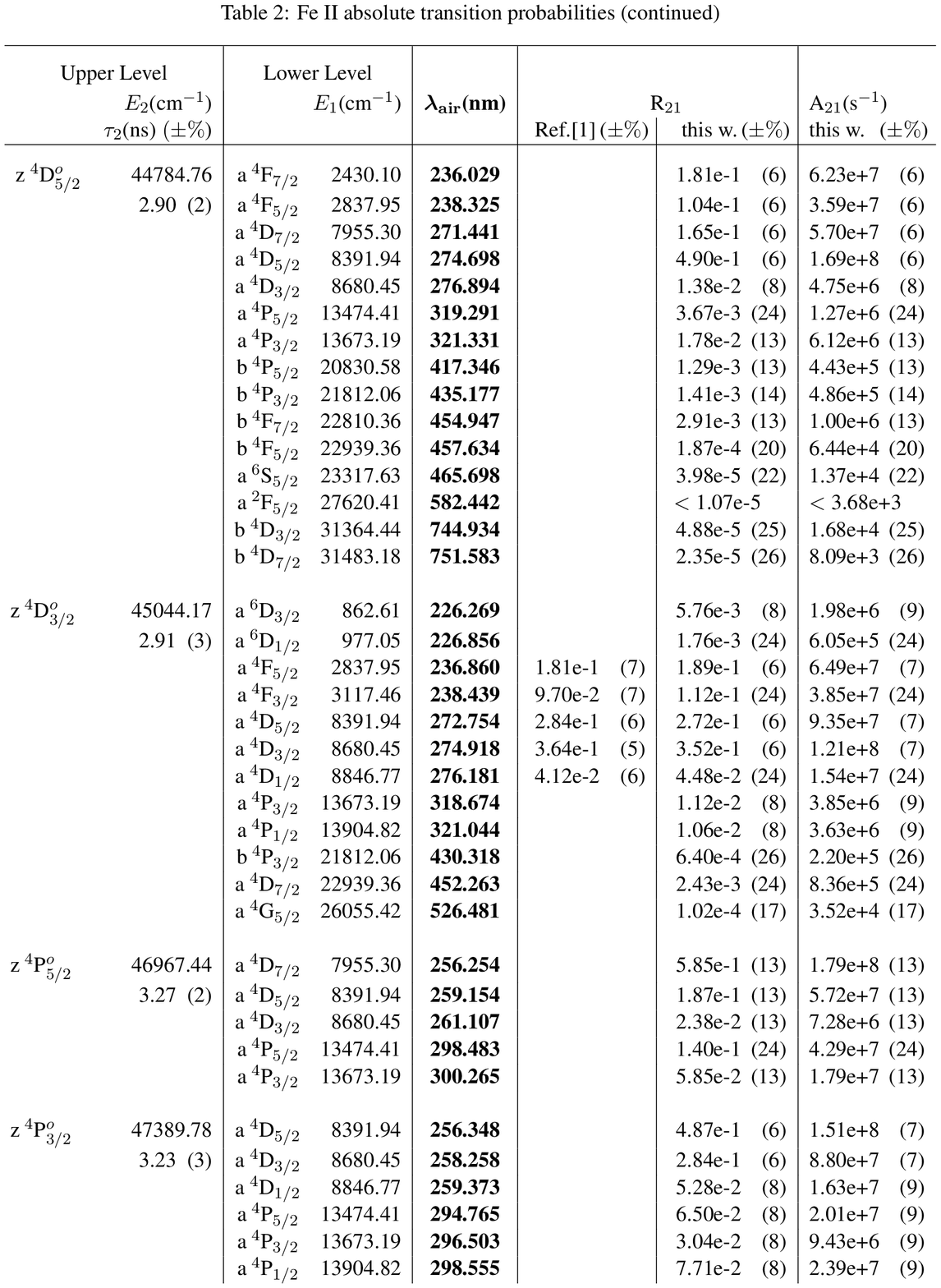}}
\vspace{3mm}\\
{[1]~\cite{BMWLLJ96}.}\\
\end{center}
\end{figure*}

\section{Acknowledgment}

This work was partly supported by the TMR Programme ``Access to
Large Scale Facilities'', Contract No. ERBFMGECT950020 (DG12) and
by Deutsche Forschungsgemeinschaft (DFG). We thank the colleagues
from Lund Laser Centre (LLC) at Lund University for hospitality
and the possibility to use their laser system.


\begin{thebibliography}{}

\bibitem[Bergeson et~al., 1996]{BMWLLJ96}
Bergeson, S.~D., Mullman, K.~L., Wickliffe, M.~E., Lawler, J.~E.,
Litzen, U.,
  and Johansson, S. (1996).
\newblock Branching fractions and oscillator strength for {Fe$\!$~II}
  transitions from the {3d$^6$($^5$D)4p} subconfiguration.
\newblock {\em Astrophys. J.}, 464:1044--1049.

\bibitem[Bi\'emont et~al., 1991]{BBKAP91}
Bi\'emont, E., Baudoux, M., Kurucz, R.~L., Ansbacher, W., and
Pinnington, E.~H.
  (1991).
\newblock The solar abundance of iron: a {"}final{"} word.
\newblock {\em Astron. Astrophys.}, 249:539--544.

\bibitem[Guo et~al., 1992]{GAPJB92}
Guo, B., Ansbacher, W., Pinnington, E.~H., Ji, Q., and Berends,
R.~W. (1992).
\newblock Beam-laser lifetime measurements for low-lying quartet states in {Fe
  $\!$II}.
\newblock {\em Phys.~Rev.~A}, 46:641--644.

\bibitem[Hannaford and Lowe, 1983]{HLo83}
Hannaford, P. and Lowe, R.~M. (1983).
\newblock Determination of atomic lifetimes using laser-induced fluorescence
  from sputtered metal vapor.
\newblock {\em Optical Engineering}, 22:532--544.

\bibitem[Hannaford et~al., 1992]{HLGN92}
Hannaford, P., Lowe, R.~M., Grevesse, N., and Noels, A. (1992).
\newblock Lifetimes in {Fe$\!$ II} and the solar abundance of iron.
\newblock {\em Astron. Astrophys.}, 259:301--306.

\bibitem[Heise and Kock, 1990]{HKo90}
Heise, C. and Kock, M. (1990).
\newblock Oscillator strengths of some weak {Fe~$\!$II} lines of astrophysical
  interest.
\newblock {\em Astron. Astrophys.}, 230:244--247.

\bibitem[Huber and Sandeman, 1986]{HSa86}
Huber, M.~C.~E. and Sandeman, R.~S. (1986).
\newblock {\em Rep. Prog. Phys.}, 49:397.

\bibitem[Kock, 1996]{Koc96}
Kock, M. (1996).
\newblock Atomic oscillator strengths from emission measurements: achievements
  and limitations.
\newblock {\em Physica Scripta}, T65:43--47.

\bibitem[Kroll, 1985]{Kroll85}
Kroll, S. (1985).
\newblock {\em Experimentelle Bestimmung von {\"U}bergangswahrscheinlichkeiten
  des einfach ionisierten Eisens}.
\newblock PhD thesis, Universit{\"a}t Hannover.

\bibitem[Kroll and Kock, 1987]{KKo87}
Kroll, S. and Kock, M. (1987).
\newblock {Fe~$\!$II} oscillator strengths.
\newblock {\em Astron. Astrophys. Suppl. Ser.}, 67:225--235.

\bibitem[Kurucz and Bell, 1995]{KBe95}
Kurucz, R.~L. and Bell, B. (1995).
\newblock Atomic spectral line data base.
\newblock {\em CD--ROM~23, Harvard Smithsonian Center for Astrophysics, April
  15}.

\bibitem[Li et~al., 1999a]{LLSJ99}
Li, Z.~S., Lundberg, H., Sikstr{\"o}m, C.~M., and Johansson, S.
(1999a).
\newblock The ferrum project: radiative lifetimes of intermediate-excitation
  states of fe~ii measured in a fluorescence signal induced by laser pumping
  from a metastable state.
\newblock {\em Eur.~Phys.~J.~D}, 6:9--12.

\bibitem[Li et~al., 1999b]{LNPWSDB99}
Li, Z.~S., Norin, J., Persson, A., Wahlstr{\"o}m, C.~G., Svanberg,
S., Doidge,
  P.~S., and Bi\'emont, E. (1999b).
\newblock Radiative properties of neutral germanium obtained from excited-state
  lifetime and branching-ratio measurements and comparison with theoretical
  calculations.
\newblock {\em Phys.~Rev.~A}, 60:198--208.

\bibitem[Norin, 1998]{Norin98}
Norin, J. (1998).
\newblock Development of a laser-pulse compression device based on stimulated
  brillouin scattering.
\newblock Master's thesis, Faculty of Technology at Lund University.

\bibitem[Raassen and Uylings, 1998a]{RUy98b}
Raassen, A.~J.~J. and Uylings, P.~H.~M. (1998a).
\newblock Critical evaluation of calculated and experimental transition
  probabilities and lifetimes for single ionized iron group elements.
\newblock {\em J. Phys. B}, 31:3137--3146.

\bibitem[Raassen and Uylings, 1998b]{RUy98a}
Raassen, A.~J.~J. and Uylings, P.~H.~M. (1998b).
\newblock On the determination of the solar iron abundance using {Fe $\!$II}
  lines.
\newblock {\em Astron. Astrophys.}, 340:300--304.

\bibitem[Schade et~al., 1988]{SMH88}
Schade, W., Mundt, B., and Helbig, V. (1988).
\newblock Radiative lifetimes of {Fe $\!$II} levels.
\newblock {\em J. Phys. B}, 21:2691--2696.

\bibitem[Schnabel and Kock, 1997]{SKo97}
Schnabel, R. and Kock, M. (1997).
\newblock Radiative lifetimes of excited {W~$\!$I} levels.
\newblock {\em Z.~Phys.~D}, 41:31--34.

\bibitem[Schnabel and Kock, 2000a]{SKo00a}
Schnabel, R. and Kock, M. (2000a).
\newblock Sub-nanosecond time-resolved nonlinear {LIF} technique used for an
  accurate f-value measurement of the {Be$\!$~I} resonance line.
\newblock {\em Phys.~Rev.~A}, 61:062506.

\bibitem[Schnabel and Kock, 2000b]{SKo00b}
Schnabel, R. and Kock, M. (2000b).
\newblock Time--resolved nonlinear laser--induced fluorescence--technique for a
  combined lifetime and branching fraction measurement.
\newblock {\em Phys.~Rev.~A}, 63:012519.

\bibitem[Schnabel et~al., 1999]{SKH99}
Schnabel, R., Kock, M., and Holweger, H. (1999).
\newblock Selected {Fe~$\!$II} lifetimes and f-values suitable for a solar
  abundance study.
\newblock {\em Astron. Astrophys.}, 342:610--613.

\bibitem[Schultz-Johanning et~al., 1999]{SSK99}
Schultz-Johanning, M., Schnabel, R., and Kock, M. (1999).
\newblock A linear paul trap for radiative lifetime measurements on ions.
\newblock {\em Eur.~Phys.~J.~D}, 5:341--344.

\bibitem[van Lessen et~al., 1998]{LSK98}
van Lessen, M., Schnabel, R., and Kock, M. (1998).
\newblock Population densities of {Fe $\!$I} and {Fe $\!$II} levels in an
  atomic beam from partially saturated {LIF} signals.
\newblock {\em J. Phys. B}, 31:1931--1946.

\end{thebibliography}

\end{document}